\magnification= 1200
\baselineskip=20 pt

\vfill\eject

\def\mh{m_h}
\def\mp{m_{\phi}}
\def\vphi{\langle \phi \rangle}
\def\rc{r_c}
\def\hphi{\hat{\phi}}
\def\dmu{\partial_{\mu}}
\def\dnu{\partial_{\nu}}
\def\l{\lambda}

\centerline{\bf Implications of radion-higgs couplings for high energy}
\centerline{\bf scattering in the Randall-Sundrum model}

\vskip 1 true in

\centerline{\bf Uma Mahanta}
\centerline{\bf Mehta Research Institute}
\centerline{\bf Chhatnag Road, Jhusi}
\centerline{\bf Allahabad, India}

\vskip .5 true in

\centerline{\bf Abstract}

In this report we derive the couplings of the Randall-Sundrum radion to
 the standard model higgs boson. We then use these couplings to determine
the J=0 partial wave amplitude for the process $hh\rightarrow h\phi $.
We find that at very high energies (i.e. $s\gg m_h^2, \mp^2$) the s wave 
partial wave unitarity is violated if $m_h>m_c\approx \sqrt{16\pi\vphi v}$
where $\vphi$ is the radion vev. Interestingly this bound is independent
of the radion mass to the leading order. We also consider the high energy
behaviour of the transition amplitudes for some other processes in the
RS scenario and compare them with their SM behaviour.

\vfill\eject

Recently several radical proposals based on extra dimensions have been put 
forward to explain the large hierarchy between the Planck scale and the
weak scale. Among them the Randall-Sundrum RS model [1] is most interesting
because it proposes a five dimensional world with a non-factorizable 
metric

$$ds^2=e^{-2k\pi\rc}\eta^{\mu\nu}dx_{\mu}dx_{\nu}-r_c^2d\theta^2.\eqno(1)$$

Here $\rc$ measures the size of the extra dimension which is an ${S^1\over
Z_2}$ orbifold. $x^{\mu}$ are the coordinates of the
four dimensional space-time.
 $-\pi \le \theta \le \pi$ is the coordinate of the extra
dimension with $\theta$ and $-\theta$ identified. Two branes extending in
the $x_{\mu}$ or space time direction are placed at the orbifold fixed 
points $\theta =0$ and $\theta =\pi$. k ia a mass parameter of the
order of the fundamental five dimensional Planck mass M. Randall and
Sundrum showed that any field with a mass parameter $m_0$ in the fundamental
five dimensional theory gets an effective four dimensional mass given
by $m=m_0 e^{-k\pi\rc}$. Thus for $k\rc\approx 12$ the weak scale
is generated from the Planck scale by the exponential warp factor
of the model.

In the original proposal of Randall and Sundrum the compactification
radius was determined by the vacuum expectation value (vev) of a scalar
field $T(x)$. However the modulus field $T(x)$ was massless and therefore
its vev was not stabilized by some dynamics. Goldberger and Wise [2] showed 
that by introducing a scalar field in the bulk with interactions localized
on the two branes it is possible to generate a potential for $T(x)$.
They also showed that this potential can be adjusted to yield a
minimum at $k\rc \approx 12$ without any extreme fine tuning of parameters.

In the Randall-Sundrum model the SM fileds are assumed to be localized
on the visible brane at $\theta =\pi$. Howvever the SM action is modified
due to the exponential warp factor. Small fluctuations of the modulus 
field $T(x)$ about its vev $\rc$ then gives rise to non-trivial
couplings of the modulus field with the SM fields.
It was shown in Ref.[3] that small fluctuations in the radion field
$\hphi$ couples with the SM fields on the visible brane through the
Lagrangian 
$$L_I={T{^\mu}_{\mu}\over \vphi}\hphi.\eqno(2)$$

Here $T^{\mu}_{\mu}$ is the trace of the energy momentum tensor of the
SM fields localized on the visible brane. $\hphi$ is a small fluctuation
of the radion field from its vev and is given by $\phi =fe^{-k\pi T(x)}
=\vphi +\hphi $. $\vphi =fe^{-k\pi \rc}$ is the vev of $\phi$ and f
is a mass parameter of the order of M.

In this report shall derive the couplings of the radion field with the
SM higgs field in the linearized approximation. We shall then use these
couplings to calculate the transition amplitude for the process
$hh\rightarrow h\phi$ at very high energies i.e. when $s\gg \mh^2 ,
\mp^2$. By requiring the $J=0$ partial wave amplitude for this process
satisfies the unitarity constraint we then derive an upper bound on the higgs
mass. Interestingly this unitarity bound on the higgs mass is independent
of the radion mass to leading order. The subleading terms has a dependence
on $\mp ^2\over s$ but it is only logarthmic and it vanishes as 
$s\rightarrow \infty$. The reasons behind considering the process
$hh\rightarrow h\phi$ are the following:

i) This process does not occur in the SM.

ii) The transition amplitude for this process is free from bad high
energy behaviour and leads to a unitarity bound on $\mh$ that is 
independent of $\mp$. 

iii) It does not receive any contribution from the tower of Kaluza-Klein
modes of the graviton and the stabilizing bulk scalar. The transition
amplitude for this process is therefore simple to compute and is
free from the uncertainties associated with processes that receive
contribution from the tower of Kaluza-Klein modes of the graviton.

The couplings of the radion field to the SM higgs field localized on the
brane at $\theta =\pi$ is completely determined by general covariance. 
The action
for the SM higgs field in the Randall-Sundrum model is given by 

$$S=\int d^4x \sqrt{-g_v}[{1\over 2}g^{\mu\nu}\dmu h\dnu h-V(h)
 ].\eqno(3)$$

Here $g^{\mu\nu}$ is the induced metric on the visible brane. In the 
abscence of graviton fluctuations about the background metric
 it is given by $g^{\mu\nu}=
e^{2k\pi T(x)}\eta^{\mu\nu}= ({\phi\over f})^{-2}\eta^{\mu\nu}$
where $\eta^{\mu\nu}$ is the Minkowski metric. $\sqrt{-g_v}=\sqrt{-det
(g_v)}=e^{-4k\pi T(x)}=({\phi\over f})^4$.
The tree level higgs potential $V(h)$ is given by
$$V(h)={\lambda\over 4} (h^4+4h^3v+4h^2v^2).\eqno(4)$$

The mass of the higgs calar can be determined from the above potential
and is given by $m_h^2=2\lambda v^2$. The couplings of the RS radion to the
higgs field is therefore given by 
$$S=\int d^4x[{1\over 2}\eta^{\mu\nu}({\phi\over f})^{ 2}\dmu h\dnu h
-{\lambda\over 4}({\phi\over f})^{ 4}(h^4+4h^3v+4h^2v^2)].\eqno(5)$$

Rescaling h and v according to $h\rightarrow {f\over \vphi} h$
and $v\rightarrow {f\over \vphi} v$ the canonically normalized
action becomes
$$S=\int d^4x[{1\over 2}\eta^{\mu\nu}({\phi\over \vphi})^2
 \dmu h\dnu h
-{\lambda\over 4}({\phi\over \vphi})^4   (h^4+4h^3v+4h^2v^2)].\eqno(6)$$

Expanding $\phi$ about its vev and keeping terms only up to linear order in 
$\hphi$ we get
$$\eqalignno{S &=\int d^4x [{1\over 2}\eta ^{\mu\nu}\dmu h\dnu h-
{\lambda\over 4}(h^4+4h^3v+4h^2v^2)]\cr
&+\int d^4x [\eta^{\mu\nu}\dmu h\dnu h-{\lambda\over 4}(h^4+4h^3v+4h^2v^2)]
{\hphi\over \vphi}+..&(7)\cr}$$

Since the trace of the energy momentum tensor of the higgs field is given 
by 
$$T^{\mu}_{\mu}=\eta^{\mu\nu}\dmu h\dnu h -\lambda (h^4+4h^3v+4h^2v^2).
\eqno(8)$$

we find that small fluctuations in the radion field from its vev couples 
to the higgs field on the visible brane  through the trace of
 its energy-momentum.
Using the classical equation of motion $\partial^2 h =-\lambda (h^3+
3h^2v^2+2hv^2)$ for the higgs field in the above expression for
$T^{\mu}_{\mu}$ the higgs-radion interaction can be written as
$$L_I=-[m_h^2h^2+\lambda v h^3]{\hphi\over\vphi}.\eqno(9)$$

The radion coupling to the higgs field therefore has a three point vertex 
and a four point vertex. We shall find later that the four point vertex 
actually leads to unitarity violation at high  enough energies.
We shall now use these couplings to determine the transition amplitude 
for the process $hh\rightarrow h\phi$ at very high energies i.e. when
$s\gg \mh^2 ,\mp^2$. Let $M_1$, $M_2$, $M_3$ denote the transition amplitudes
 due to s, t, u channel higgs exchanges and $M_4$ denote the transition
 amplitude for due to the four point $hhh\phi$ vertex. We then find that
 at very high energies to leading order in $\mh^2\over s$ and $\mp^2\over s$

$$M_1={6\l v\over \vphi}{\mh^2\over s-\mh^2}\approx {6\l v\over \vphi}
{\mh^2\over s}.\eqno(10a)$$

$$M_2={6\l v\over \vphi}{\mh^2\over t-\mh^2}\approx -{12\l v\over \vphi}
{\mh^2\over s}{1\over \alpha -\beta x}.\eqno(10b)$$

$$M_2={6\l v\over \vphi}{\mh^2\over u-\mh^2}\approx -{12\l v\over \vphi}
{\mh^2\over s}{1\over \alpha + \beta x}.\eqno(10c)$$

and 

$$M_4={\l v\over \vphi}.\eqno(10d)$$

where $\alpha =1-{\mh^2 +\mp^2\over s}$ and $\beta =1-{3\mh^2 +\mp^2\over s}$.
$x=\cos\theta$ where $\theta$ is the angle between the outgoing h and one 
of the incoming h. We shall consider energies far above the higgs pole
to avoid the strong rise in the transition amplitude near the pole.
It is clear that as $s\rightarrow \infty$ the terms that constitute
a potential threat to unitarity are the ones that diverge with s or
at least remains constant. We find from the above expressions that as 
$s\rightarrow \infty$, the first three transition amplitudes tend to zero.
However $M_4$ remains a constant. Although the total transition
amplitude has acceptable high energy behaviour, it can lead to unitarity
violation for sufficiently large values of $\l$ or small $\vphi$.
In other words either $\mh$ must be sufficiently small or $\vphi$
must be large enough to avoid unitarity violation.
In order to determine the unitarity bound on $\mh$ implied by
the process $hh\rightarrow h\phi$ we shall follow Lee, Quigg and
 Thacker [4] and confine our attention to the J=0 partial wave amplitude.
We find that

$$\eqalignno{a_0 & \approx{\l v\over 16\pi\vphi}
[1+ 6{\mh^2\over s} -12{\mh^2\over s\beta}\ln{\alpha +\beta\over \alpha
-\beta}]\cr
&\approx {\l v\over 16\pi \vphi}
[1+ 6{\mh^2\over s} +12{\mh^2\over s\beta}\ln {\mh^2\over s}]\cr
&\approx {\l v\over 16\pi \vphi}.&(11)\cr}$$

as ${\mh^2\over s}\rightarrow 0$. The unitarity constraint 
$\vert a_0 \vert <{1\over 2}$ 
then implies that $\mh^2 <16\pi\vphi v$. In other words the higgs mass 
must be less than $4 \sqrt {\pi \vphi v}$ so that the J=0 partial wave
amplitude does not violate unitarity. For $\vphi \approx 1$ Tev
the unitarity bound on $\mh $ is about 3.5 Tev. We would like
to note firstly that in SM the unitarity bound on $\mh$ that follows
from the 
somewhat analogous process $hh\rightarrow hh$ is given by 
$\mh <\sqrt {8\pi\over 3} v\approx .7 $ Tev which is somewhat smaller 
than the value obtained in this report. Secondly although the processes
$W^+_LW^-_L\rightarrow W^+_LW^-_L$, $W^+_LW^-_L\rightarrow Z_LZ_L$, 
$W^+_LW^-_L\rightarrow hh$ and $Z_LZ_L\rightarrow hh$ have acceptable
high energy behaviour in the SM model they exhibit bad high energy
behaviour in the context of the Randall-Sundrum scenario. The bad
high energy behaviour arises from the diagram that involves the
exchange of radion. These processes however also receive contributions
from the s channel exchange of Kaluza-Klein gravitons which 
could give rise to transition amplitudes with much worse
high energy behaviour.
 On the other hand the process $hh\rightarrow hh$
exhibits acceptable high energy behaviour both in the SM and 
the RS scenario if we neglect the contribution of KK gravitons.
 The  s wave transition amplitude for the latter process
however depends both on $m_h$ and $\mp$ in the RS secnario.
The unitarity constraint $\vert a_0 \vert <{1\over 2}$ therefore
gives a bound on $m_h$ that depends on $\mp $. This explains
the reason why we chose the process $hh\rightarrow h\phi$
which gives a unitarity bound on $m_h$ that is independent
of $\mp $.

 The four point vertex $hhh\phi$ 
in the higgs-radion interaction Lagrangian is crucial for the unitarity
bound discussed in this paper.
 The presence of this term in $T^{\mu}_{\mu}$ is easily
understood. The electro-weak symmetry breaking also breaks the conformal
symmetry. The conformal symmetry is broken not only by the higgs mass term
in the potential but also the trilinear higgs interaction term.
 The unitarity violation discussed
 in this paper signals that the higgs-radion coupling becomes strong
and non-perturbative as $\l$ approaches its unitarity limit value.
Therefore higher order radiative corrections and non-perturbative effects
 become imporatnt and they
could significantly change  the unitarity bound presented here.
The estimation of such non-perturbative corrections although important 
will not be considered in this paper.

\vskip .3 true in

\centerline{\bf References}

\item{1.} L. Randall and R. Sundrum, Phys. Rev. Lett. 83, 3370 (1999).

\item{2.} W. D. Goldberger and M. B. Wise, Phys. Rev. Lett. 83, 4922 (1999).

\item{3.} C. Csaki, M. Graesser, L. Randall and J. Terning, hep-ph/9911406;
W. D. Goldberger and M. B. Wise, hep-ph/9911457.

\item{4.} B. W. Lee, C. Quigg and H. B. Thacker, Phys. Rev. D 16 282 (1977).

\end